\documentclass[12pt]{article}
\usepackage{graphics}
\usepackage{amsthm}
\usepackage{amssymb}
\usepackage{amsmath}
\usepackage{amsfonts}
\usepackage{epic}
\usepackage{hyperref}
\usepackage{epsfig}
\usepackage{bm}
\usepackage{amscd}
\usepackage{amsmath,esint}

\theoremstyle{plain}

\theoremstyle{definition}

\def\be{\begin{equation}}
\def\ee{\end{equation}}
\smallskip

\begin{document}

\headsep=-0.5cm

\begin{titlepage}
\begin{flushright}
\end{flushright}
\begin{center}
\noindent{{\LARGE{Overflying Nilpotent Horizons}}}

\smallskip
\smallskip
\smallskip

\smallskip
\smallskip

\smallskip

\smallskip

\smallskip
\smallskip
\noindent{\large{Jos\'e Figueroa$^{1,2}$, Gaston Giribet$^{3}$, Anibal Neira-Gallegos$^{1}$, Julio Oliva$^{1}$, Marcelo Oyarzo$^{1,4}$}}
\end{center}

\smallskip

\smallskip
\smallskip

\centerline{$^{1}$Departmento de F\'{\i}sica, Universidad de Concepci\'on}
\centerline{{\it Casilla 160-C, Concepci\'on, Chile.}}
\smallskip
\centerline{$^{2}$Physique Mathématique des Interactions Fondamentales, Université Libre de Bruxelles,}
\centerline{{\it Campus
Plaine - CP 231, 1050 Bruxelles, Belgium.}}

\smallskip
\centerline{$^{3}$Department of Physics, New York University}
\centerline{{\it 726 Broadway, New York, NY 10003, USA.}}
\smallskip
\centerline{$^4$Department of Applied Science and Technology, Politecnico di Torino,}
\centerline{{\it C.so Duca degli Abruzzi, 24, I-10129 Torino, Italy.}}

\smallskip
\smallskip

\smallskip
\smallskip

\smallskip

\smallskip

\begin{abstract}
We study solutions of Einstein equations with negative cosmological constant in five dimensions that describe black holes whose event horizons are homogeneous, anisotropic spaces. We focus on the case where the constant-time slices of the horizon are the Nil geometry, the Thurston geometry associated to the Heisenberg group. For such spaces, we analyze the symmetries both in the asymptotic region and in the near horizon region. We compute the associated conserved charges, which turn out to be finite and admit a sensible physical interpretation. We analyze the thermodynamics of the Nil black hole, and we present a stationary spinning generalization of it in the slowly rotating approximation. 
\end{abstract}

\end{titlepage}

\newpage


\section{Introduction}

One of the most interesting aspects of gravity theory in five or more dimensions is the existence of isolated event horizons with varied geometries, including spaces with nontrivial topology, spaces of non-constant curvature, disconnected horizons, among others. The most significant example is undoubtedly the black ring solution \cite{blackring1, blackring2}, along with its generalizations \cite{blacksaturn}. There also exist generalizations of topological black holes \cite{topological}, whose simplest versions have base manifolds composed of the direct product of homogeneous spaces. In five dimensions, there are also black hole solutions whose event horizons have constant-time sections given by other types of spaces, including non-homogeneous, non-Einstein spaces. Among these, there are black holes whose constant-time horizon foliations are anisotropic spaces that correspond to some of Thurston's geometries \cite{Nil, Nil2, Nil3}; i.e. the eight geometries that appear in the geometrization problem of topological spaces in three dimension. These solutions exhibit a richer geometric structure than topological black holes, as well as an asymptotic behavior that differ from those of maximally symmetric spacetimes. Due to the latter, studying physical properties of such black holes, such as their conserved charges or thermodynamic variables, requires a general framework that works for a wide class of spacetimes. In this paper, we will address this problem. We will work with Einstein theory in five dimensions with negative cosmological constant $\Lambda=-1/\ell^2$. This is defined by the Einstein-Hilbert action
\begin{equation}
I= \frac{1}{16\pi G}\int d^5x \,\sqrt{-g}\left(\, R+\frac{2}{\ell^2}\,\right) \, + \, I_{B}\label{Einstein}
\end{equation}
where $I_{B}$ stands for boundary terms. This theory admits black hole solutions with diverse horizon structures. We will consider as a working example the case of black holes with horizons given by the so-called Nil space \cite{Nil}, the geometry associated to the Heisenberg group. For a solution of this type, we will extend the method of computing conserved charges associated with the infinite-dimensional symmetries that emerge near the horizons. This method was recently employed in \cite{Juan} for the analysis of conserved charges and thermodynamic variables of the Black Ring. Here, we will adapt the computation to the case of black holes with Nil horizons. We will analyze their asymptotic symmetries both in the region near the horizon and in the asymptotic region, compute the associated charges, and the thermodynamic variables of the solution.

The paper is organized as follows: in section 2, we introduce the geometry of black holes with Nil horizon. In section 3, we will study the symmetries of these spacetimes, both in the asymptotic far region and in the near horizon region. In section 4, we will study the thermodynamics of these solutions, which amounts to compute the near horizon charges. We will conclude in section 5 by presenting a spinning generalization of the solution valid in the slowly rotating approximation.

\section{Nilpotent black holes}

Let us consider the black hole solution found in \cite{Nil}, whose metric is
\begin{equation}
ds^2=-V(r)\,
dt^{2}+\frac{dr^2}{V(r)}
+r^{4/3}\left(\, dx^{2}+dy^{2}\right) +r^{8/3}\Big( \,dz-\frac{2}{3\ell} \,x\,dy\Big) ^{2}\ .
\end{equation}
with the function
\begin{equation}
V(r)= \frac{2 }{11\ell^2}r^{2}-\frac{2\mu}{r^{5/3}}\label{potential}
\end{equation}
and where $t\in \mathbb{R}$, $r\in \mathbb{R}_+$. $\mu$ is an integration constant which we will take to be positive. As we will see, $\mu$ turns out to be proportional to the mass of the solution. We will denote $x^1=x$, $x^2=y$, $x^3=z$. Each of the space-like coordinates $x^A$ on the constant-$t$, constant-$r$ base manifold covers a domain $D_A$ ($A=1,2,3$). A compact base manifold is possible provided one consider a quotient by a discrete subgroup of its symmetry group. Locally, the base manifold is a fibration of the Nil geometry. In fact, if we define the rescaled coordinates $\hat{x}=(2/3\ell)^{\frac 12}x$, $\hat{y}=(2/3\ell)^{\frac 12}y$, $\hat{z}=z$ and evaluate at $r=(3\ell /2)^{\frac 43}$, we find the Nil metric
\begin{equation}
ds^2_{\text{Nil}} = \frac{9\ell^2}{4}\, \Big(d\hat{x}^{2}+d\hat{y}^{2}+(d\hat{z}-\hat{x}d\hat{y})^2\Big)\, .
\end{equation}

Nil manifold is one of the eight Thurston geometries. It fibers the 2-dimensional Euclidean space producing a twisted product $\mathbb{R}^2\times \mathbb{R}$. It is the geometry of the Heisenberg group, and hence the name Nil, for ``Nilpotent''. It can be associated to other spaces relevant in physics: it is a section of the Bianchi type II solution to 4-dimensional Einstein equations. Since not only the base manifold but all constant-$t$, constant-$r$ slices of the space (\ref{lametrica}) are locally Nil, we will refer to this solution as the Nil black hole. 

The metric describes a black hole whose event horizon is located at $r_H=(11\ell^2\mu )^{\frac{3}{11}}$. The horizon is regular; however, the solution exhibits a curvature singularity at the origin $r=0$, where the Kretschmann scalar is found to diverge as $R_{\mu\nu\sigma\eta}R^{\mu\nu\sigma\eta}\sim \mu/r^{\frac{22}{3}}$.

In order to study the properties of the solution (\ref{lametrica}), such as its asymptotic behavior and the conserved charges associated to it, it is convenient to redefine the radial coordinate as $\rho =r^{1/3}$. This yields
\begin{equation}
ds^2=-\rho ^4\,\left( \frac{2 }{11\ell^2}\rho ^{2}-\frac{2\mu}{\rho ^{9}}%
\right) dt^{2}+\frac{9\, d\rho ^{2}}{\left( \frac{2 }{11\ell^2}\rho ^{2}-\frac{2\mu}{\rho ^{9}}%
\right)}+\rho ^{4}\Big( dx^{2}+dy^{2}\Big) +\rho ^{8}\left( dz-\frac{2}{3\ell}\, x\,dy\, \right) ^{2}\ .\label{lametrica}
\end{equation}%
The ground state corresponds to the particular case $\mu=0$, which we will consider as the background configuration with respect to which we compute the conserved charges. In the asymptotic region, spacetime (\ref{lametrica}) exhibits anisotropic scale invariance, which we will discuss below. This makes it to be related to the solutions discussed in the literature in the context of non-relativistic holography \cite{Lifshitz}. In particular, it is easy to see that, if we define the new radial coordinate $\hat{r}=\rho^2$ and rescale $t$ and $z$ accordingly, the $dy=dz=0$ slices of the manifold describe a three-dimensional Lifshitz black hole, which asymptotes to $ds^2_{\text{Lif}}\simeq -\hat{r}^{2(\text{z}+1)}dt^2+\hat{r}^{-2}d\hat{r}^{2}+\hat{r}^2dz^2$ at large $\hat{r}$, with dynamical exponent $\text{z}=\frac 12$ (where we fixed $\ell^2={8}/{99})$. Similarly, the $dx=dy=0$ slices are diffeomorphic to a three-dimensional Lifshitz black hole with negative dynamical exponent $\text{z}=-\frac 14$. We will discuss the asymptotic scale invariance below. 

\section{Symmetries and Noether charges}

The local isometry group of the Nil black hole solution has dimension 5, and in the case $\mu=0$ gets enhanced to a 6-dimensional group generated by the Killing vectors
\begin{eqnarray}
\xi _{1} =-3\ell \partial _{y}\ , \ \ \ \ \
\xi _{2} =3\ell \partial _{x}+2y\partial _{z}\ , 
\end{eqnarray}
together with anisotropic scale transformation
\begin{eqnarray}
\xi _{3} &=&3t\partial _{t}-\rho \partial _{\rho }+2x\partial
_{x}+2y\partial _{y}+4z\partial _{z}\ , 
\end{eqnarray}
the spacetime translations
\begin{eqnarray}
\xi _{4} =\partial _{z}\ , \  \ \ \
\xi _{5} =\partial _{t}\, ,
\end{eqnarray}
and the additional special transformation 
\begin{equation}
\xi_6= y\partial_x - x\partial_y-\frac{1}{3\ell }(x^2-y^2)\partial_z\, .
\end{equation}
These Killing vectors obey the following Lie algebra:
\begin{eqnarray}
[\xi_{1}, \xi_{2}] = -6\xi_{4}\, , \ \ \ \ \label{Heisenberg} 
\end{eqnarray}
together with
\begin{eqnarray}
[\xi_{1}, \xi_{3}] = 2\xi_{1}\, , \ \ \ \ [\xi_{2}, \xi_{3}] = 2\xi_{2}\, , \ \ \ \
[\xi_{3}, \xi_{4}] = -4\xi_{4}\, , \ \ \ \ [\xi_{3}, \xi_{5}] = -3\xi_{5}\, \label{propers}
\end{eqnarray}
and
\begin{eqnarray}
[\xi_{1}, \xi_{6}] = -\xi_{2}\, , \ \ \ \ [\xi_{2}, \xi_{6}] = \xi_{1}\;  \label{semisimple}
\end{eqnarray}
with the Lie products that are not shown here being zero. It might be convenient to rescale some generators, e.g. $\xi_{4}\to 6\xi_{4}$, to write the algebra in a more familiar way. The isometries are globally well defined for $x, y, z\in \mathbb{R}$, i.e. $\cup_{A=1}^3D_A=\mathbb{R}^3$. 

The algebra above contains non-trivial Abelian ideals, e.g. the one generated by $\{\xi_{4}, \xi_{5}\}$. It also contains non-Abelian ideals. As said, the symmetry generated by $\xi_{3}$ breaks down in the case $\mu\neq 0$, while the other five isometries generated by $\{\xi_1, \xi_2,\xi_4,\xi_5, \xi_6\}$ remain unbroken. In the massive case ($\mu> 0$) the non-trivial part of the algebra reduces to the nilpotent Heisenberg algebra (\ref{Heisenberg}), with $\xi_4$ being a central element, in direct sum with the semisimple algebra (\ref{semisimple}), and both of them in direct sum with $\xi_5$. In the massless case ($\mu =0$) the algebra contains additional non-Abelian dimension-2 nilpotent proper subalgebras (\ref{propers}). The six isometries persist asymptotically, at large $\rho $ for all $\mu$. The Killing vector $\xi _{3}$ generates an anisotropic scale transformation 
\begin{eqnarray}
t\to \lambda^{3}t\, , \ \ \ \ \rho\to \lambda^{-1}\rho \,
, \ \ \ \
x\to \lambda^{2}x\, , \ \ \ \ y\to \lambda^{2}y
\, , \ \ \ \ z\to \lambda^{4}z \, ,
\end{eqnarray}
similar to those appearing in the Lifshitz spacetimes \cite{Lifshitz} and their generalizations. Notice that we can consistently assign length dimensions to the coordinates as follows: if we denote length dimensionality as $[\ell ]=1$, and so $[G]=3$, then we have $[t]=1$, $[x]=[y]=\frac 13$, $[z]=-\frac 13$, $[\rho ]=\frac 13$.

The Nil black hole solution can be accommodated in the following asymptotic condition at infinity,
\begin{eqnarray}
g_{tt} &=&-\frac{2 }{11\ell^2}\,\rho ^{6}+\mathcal{O}\left( \rho ^{-5}\right)\,, \ \ \ \ \ \ \ \ \ \ \ \ \
g_{\rho \rho } =\frac{99\ell^2}{2 }\, \rho ^{-2}+\mathcal{O}\left( \rho
^{-13}\right) \,,\label{Losunos}\\
g_{11} &=&\rho ^{4}+\mathcal{O}\left( \rho
^{-7}\right) \,, \ \ \ \ \ \ \ \ \ \ \ \ \ \ \ \ \ \ \ \ \ \ 
g_{12} =\mathcal{O}\left( \rho ^{-2}\right)\,,
\\
g_{13} &=&\mathcal{O}\left( \rho
^{-7}\right) \,, \ \ \ \ \ \ \ \ \ \ \ \ \ \ \ \  \ \ \ \ \ \ \  \ \ \ \ \ 
g_{22} =\frac{1 }{9\ell^2}x^{2}\rho ^{8}+\rho ^{4}+\mathcal{O}\left( \rho
^{-7}\right) \,,\\
g_{23} &=&-{\frac{2 }{3\ell }}\,x\rho ^{8}+\mathcal{O}\left( \rho
^{-7}\right) \,, \ \ \ \ \ \ \ \ \ \ \ \ \
g_{33} =\rho ^{8}+\mathcal{O}\left( \rho ^{-7}\right)\,,\label{Losotros}
\end{eqnarray}
with the gauge fixing conditions $g_{\rho t}=g_{\rho A}=0$, and with the rest of the components being of order $\sim \mathcal{O}({\rho }^{-5})$. One can relax these conditions to gather contributions $g_{1A}\sim \mathcal{O}(\rho^8)$; in fact, there exists a way of expressing the falling off conditions above that makes the symmetry in the $(x,y)$ plane manifest; see (\ref{Losunosno})-(\ref{Losotrosno}) below. Notice that in (\ref{Losunos})-(\ref{Losotros}) we are using the notation $g_{AB}$ to label the components on the base manifold, e.g. $g_{23}=g_{yz}=g_{zy}$. These conditions yield finite conserved charges. It can also be shown that the symmetry group that preserves such large-$\rho $ behavior is generated by a finite-dimensional Lie group which coincides with the exact isometry group of the background solution $\mu=0$.

The mass of the Nil black hole (\ref{lametrica}) can be computed by different methods, the most efficient one being the phase space method. This yields a result proportional to $\mu$; more precisely,
\begin{equation}
M=\frac{\text{Vol}_3\, \mu }{3\pi G}\ \ , \  \ \ \ \ \ \text{Vol}_3 =\int_{\cup_A D_A} d^3x .\label{LaMasa}
\end{equation}
In the case the domains of integration of the coordinates $x^A$ are $D_A\in \mathbb{R}$, the charges diverge due to the non-compactness of the base manifold, and so one has to interpret all physical quantities as densities per unit of 3-volume. Notice that $\text{Vol}_3$ has length dimension $\frac 13$. Below, we will discuss this value of the mass in relation to the thermodynamics.

\section{Thermodynamics}

The most efficient method to derive the thermodynamic variables of a solution like (\ref{lametrica}) is the near horizon computation. Near an isolated horizon we can always consider the following expansion in powers of $\eta $
\begin{equation}\label{Desarrolloss}
    g_{\mu \nu} = g_{\mu \nu}^{(0)}\, +\, g_{\mu \nu}^{(1)} \, \eta \, +\,  g_{\mu \nu}^{(2)} \, \eta^2 \, +\, \mathcal O(\eta^3)
\end{equation}
where $\eta$ measures the distance from the horizon: the horizon location is $\eta =0$. $g_{\mu \nu}^{(n)}$ ($n\in\mathbb{Z}_{\geq 0}$) are functions of $v$ and $x^A$ ($A=1,2,3$), and are independent of $\eta $. Next, we impose the following boundary conditions
\begin{equation}
g^{(0)}_{vv}=0\, , \ \ \ g^{(1)}_{vv}=-2\kappa \, , \ \ \  g^{(0)}_{vA}=0, 
\end{equation}
together with the gauge condition 
\begin{equation}\label{Desarrolloss2}
g_{\eta v}=-1 \, , \ \ \   g_{\eta \eta}=0 \, , \ \ \   g_{\eta A}=0\, ; 
\end{equation}
$\kappa $ is the surface gravity. Boundary conditions (\ref{Desarrolloss})-(\ref{Desarrolloss2}) are the near horizon form studied in \cite{Donnay1, Donnay2} and they were considered in many different scenarios. It was shown in \cite{Donnay2} that the asymptotic Killing vectors preserving the asymptotic conditions (\ref{Desarrolloss})-(\ref{Desarrolloss2}) form an infinite-dimensional algebra that includes supertranslations and superrotations, cf. \cite{Strominger}. The horizon supertranslation \cite{Hawking} symmetry corresponds to a local shift in the null coordinate on the horizon, namely $v\to v+f(x,y,z)$. These are generated by
\begin{equation}
\hat{\xi}_5 \, =\, P(x^A)\,\partial_{v}\,  \label{supert}
\end{equation}
which gives an infinite dimensional extension of $\xi_5$. 

The Noether charges associated to the near horizon supertranslations (\ref{supert}) are given by \cite{Donnay2}
\begin{equation}
	Q[P\partial_{v}] = \frac{\kappa }{8 \pi G} \int_{\bar{H}_+} d^3x\,\sqrt{\text{det}{g_{\text{AB}}^{(0)}}}\,P(x^C)\,,
  \label{eq_cargas_Barnich}
\end{equation}
where, again, we have used Latin indices $A, B=1,2,3$ to denote the space-like coordinates on the base manifold. The charges are calculated as integrals over constant-$v$ slices on the horizon, which we denote $\bar{H}_+$. The Dirac brackets of these charges yield the same algebra,
\begin{equation}
\{Q[P_1\partial_{v}],Q[P_2\partial_{v}]\}=0,
\end{equation}
and do not pick up any central extension.

In order to analyze the near horizon geometry of the Nil black hole, we define the Eddington type coordinates
\begin{equation}
v=t-\int^r\frac{d\text{r}}{V(\text{r})} \, , \ \ \ \ \ \ \eta = r-r_H\, ,
\end{equation}
This yields
\begin{eqnarray}
g^{(0)}_{11}= \rho^4_H \, , \ \ \ g^{(0)}_{22}= \rho^4_H+\frac{4}{9\ell^2}\rho^8_H x^2\, , \ \ \  g^{(0)}_{23}=g^{(0)}_{32}= -\frac{2}{3\ell}\rho_H^8x\, , \ \ \  g^{(0)}_{33}= \rho^8_H\, , 
\end{eqnarray}
with the components $g^{(0)}_{\mu \nu }$ omitted here being zero, and with $\rho^{11}_H=11\ell \mu$. From this, we have ${\text{det}g_{AB}^{(0)}} = \rho^{16}_H$ or, more generally,
\begin{equation}
\sqrt{\text{det}g_{AB}} \, = \, \rho^8\, .\label{det}
\end{equation}

It is worth noticing that, under transformation (\ref{supert}), the metric (\ref{lametrica}) picks up a non-diagonal contribution
\begin{equation}
\delta g^{(0)}_{Av } =0 \, , \ \ \ \ \ \ \delta g^{(1)}_{Av } = -2\kappa \partial_{A}P(x^B)
\end{equation}
which yields terms of order $g_{Av }\sim \mathcal{O}(\eta )$ that vanish sufficiently fast near the horizon $\eta = 0$. 

The Wald entropy corresponds to the charge
\begin{equation}
Q[\partial_v]= \frac{\kappa}{2\pi }\frac{A}{4G}\,,\label{Waldo}
\end{equation}
with the area $A=\rho^{8}_H\text{Vol}_3$ being the integral on the constant-$v$ slices of the horizon at $\rho=\rho_H$ ($\eta = 0$). $\kappa $ is the surface gravity: defining radial coordinates $\sigma^2={2}{\kappa^{-1}} \eta $ and expanding the metric near the horizon, we get the Rindler space with Unruh temperature $\kappa /2\pi$ in direct product with the fibration of the Nil geometry, namely $
ds^2\simeq -(\kappa^2 \sigma^2dt^2-d\sigma^2) +ds^2_{\text{Nil}}+...$, up to terms $\sim \mathcal{O}(\sigma^4)$. With all this, we can evaluate the charge (\ref{Waldo}), which gives the Hawking temperature and the Bekenstein-Hawking entropy of the Nil black hole. This yields
\begin{equation}
T=\frac{\kappa}{2\pi } = \frac{\rho_H^3}{6\pi \ell^2} \, , \ \ \ \ \ \ S=\frac{\rho_H^8 \text{Vol}_3}{4G}\, ,\label{QT}
\end{equation}
respectively. The mass of the black hole is given by
\begin{equation}
M=\frac{\rho^{11}_H\text{Vol}_3}{33\pi G \ell^2}\, .\label{Masita}
\end{equation}
which agrees with (\ref{LaMasa}). This result for the mass is positive definite --something not entirely obvious in the case of solutions with event horizons of non-trivial topology-- and is shown to obey the first law of black hole mechanics, $dM=T\,dS$. 

Equation (\ref{det}) shows that the area of the hyper-spheres of constant-$\rho$ at fixed $t$ grow as $\sim \rho^{8}$, while the Newtonian-like term in the metric (\ref{lametrica}) is $\sim 2\mu/\rho^{9}$. This is related to the fact that the potential function (\ref{potential}) can be written as 
\begin{equation}
V(r)=-\frac{6\pi G}{r^{5/3}\text{Vol}_3}\left(\,{M}+\frac 89\, {M}_{\Lambda }(r)\,\right) \  \ \ \text{with} \  \ \ {M}_{\Lambda }(r)=\frac{\Lambda}{8\pi G}\int_{\cup_i D_i}d^3x\int^{r}_{0} dr\, \sqrt{\text{det}g_{AB}^{(0)}} .\nonumber
\end{equation}
That is, the constant-density contribution of the cosmological constant $\Lambda$ can be understood as a gravitational energy contribution to the effective mass enclosed in a volume of radius $r$ weighted with a geometrical Gauss factor $\frac 89$.

The expression for the mass (\ref{Masita}) satisfies the Smarr type formula $TS=\frac{11}{8}M$. This implies that the Nil black hole has negative free energy $F=M-TS=-\frac 38 M<0$ for all values of the mass, with positive entropy $S=-\frac{\partial F}{\partial T}>0$ and positive specific heat $c=\frac{\partial M}{\partial T}>0$. This implies that the Nil black holes may be in thermal equilibrium with their own Hawking radiation; i.e. they are always {\it large} black holes, regardless their mass relative to the spacetime curvature.

\section{Introducing spin}

Before concluding, let us present a spinning generalization of the metric (\ref{lametrica}), which we derive in the slowly rotating approximation. In fact, it can be shown that the following metric satisfies Einstein equations in vacuum for small values of $\delta a $
\begin{eqnarray}
ds^2 &=&-\rho^{4}\left( \frac{2 }{11\ell^2}\rho ^{2}-\frac{2\mu }{\rho ^{9}}%
\right) dt^{2}+{9\,\left(\frac{2 }{11\ell^2}\rho ^{2}-\frac{2\mu }{\rho ^{9}}\right)^{-1} d\rho ^{2}}{}
+\rho ^{4}\left( dx^{2}+dy^{2}\right)
\nonumber \\
&&\, +\rho ^{8}\left( dz+{\frac{1}{3\ell}}\left(
ydx-xdy\right) \right) ^{2} + \frac{\delta a}{\rho ^{5}}\,dt\left( dz+\frac{1 }{3\ell }\left(
ydx-xdy\right) \right)\, .\label{larotante}
\end{eqnarray}%
This metric solves Einstein equation at first order in the parameter $\delta a$. Metric (\ref{larotante}) satisfies the large $\rho $ expansion (\ref{Losunos})-(\ref{Losotros}). $\delta a$ in the last term of (\ref{larotante}) controls the rotational dragging; notice that the metric is treating now the coordinates $x$ and $y$ in equal footing, which is achieved by performing the change of coordinates $z\to z-xy/(3\ell )$, which induces the following change in the asymptotic Killing vectors: 
\begin{equation}
\xi_1\to \xi _{1} =-3\ell \partial _{y}+x\partial _{z}\, , \ \ \ \ \ \xi_2\to \xi _{2} =3\ell \partial _{x}+y\partial _{z}\, , \ \ \ \ \ \xi_6\to \xi_6=y\partial_x -x\partial_y\, ,
\end{equation}
the latter being the rotation in the $(x,y)$ plane. In this frame, the asymptotic conditions at infinity take the form
\begin{eqnarray}
g_{tt} &=&-\frac{2 }{11\ell^2}\,\rho ^{6}+\mathcal{O}\left( \rho ^{-5}\right)\,, \ \ \ \ \ \ \ \ \ \ \ \ \
g_{\rho \rho } =\frac{99\ell^2}{2 }\, \rho ^{-2}+\mathcal{O}\left( \rho
^{-13}\right) \,,\label{Losunosno}\\
g_{11} &=&\frac{1 }{9\ell^2}\,y^{2}\rho ^{8}+\rho ^{4}+\mathcal{O}\left( \rho
^{-7}\right) \,, \ \ \ \ \ \ \ 
g_{12} =-\frac{1 }{9\ell^ 2}\,xy\rho ^{8}+\mathcal{O}\left( \rho ^{-2}\right)\,,
\\
g_{13} &=&{\frac{1 }{3\ell }}\,y\rho ^{8}+\mathcal{O}\left( \rho
^{-7}\right) \,, \ \ \ \ \ \ \ \ \ \ \ \ \ \ \ \ 
g_{22} =\frac{1 }{9\ell^2}x^{2}\rho ^{8}+\rho ^{4}+\mathcal{O}\left( \rho
^{-7}\right) \,,\\
g_{23} &=&-{\frac{1 }{3\ell }}\,x\rho ^{8}+\mathcal{O}\left( \rho
^{-7}\right) \,, \ \ \ \ \ \ \ \ \ \ \ \ \
g_{33} =\rho ^{8}+\mathcal{O}\left( \rho ^{-7}\right)\,,\label{Losotrosno}
\end{eqnarray}
which makes the rotational symmetry in the ($x,y$) plane manifest; cf. (\ref{Losunos})-(\ref{Losotros}). The metric also obeys the requirement
\begin{equation}
g_{tA}\,\simeq \,\mathcal{O}(\rho^{-5})\, .
\end{equation}

The angular momentum of the solution results proportional to $\delta a$. More precisely, the Noether charge associated to the Killing vector $\xi_{4}$ gives
\begin{equation}
Q[\xi_4] = \frac{13 \,\delta a\, \text{Vol}_3}{96\pi G}\, ,\label{Q1}
\end{equation}
which corresponds to the density of momentum along the $z$ direction; notice that, indeed, (\ref{Q1}) has length dimension $\frac 13$. On the other hand, the charge associated to the Killing vector $\xi_6$ is 
\begin{equation}
Q[\xi_6] = \frac{13 \,\delta a\, {I}}{288\pi G\ell}\, ,  \  \ \ \ \ \ {I}=\int_{\cup_A D_A}d^3x\,(x^2+y^2)\label{Q2}
\end{equation}
which depends on the domains of integration of the variables on the base manifold. (\ref{Q2}) is actually dimensionless and corresponds to the angular momentum. The integral $I$ can be thought of as the moment of inertia associated to rotations on the ($x,y$) plane, which is exactly what $\xi _6=y\partial_x - x\partial_y$ generates. This invites to use polar coordinates $x=\text{r}\cos{\varphi}$, $y=\text{r}\sin{\varphi}$, with $\xi_6=-2\partial_{\varphi}$. In these coordinates, the $dt=d\rho =0$ sections of the metric take the form $ds^2_{\text{Nil}}=\rho^4(d\text{r}^2+\text{r}^2d\varphi^2 )+\rho^8 (dz-\frac{1}{3\ell}\text{r}^2d\varphi )^2$.

\section{Conclusions}

In this paper, we studied solutions of Einstein equations with negative cosmological constant in five dimensions which describe black holes whose event horizons are homogeneous but anisotropic spaces. We focused on the case in which the constant-time slices of the horizon is the Nil geometry, the Thurston geometry associated to the Heisenberg group. For such spaces, we analyzed the symmetries both in the asymptotic region and in the near horizon region. We computed the associated conserved charges, which turned out to be finite and lead to a sensible physical interpretation; see (\ref{Masita}). We analyzed the thermodynamic variables (\ref{QT}) of these black holes, and we derived a stationary spinning generalization of it in the slowly rotating approximation, which is given by (\ref{larotante}) and carries angular momenta (\ref{Q1})-(\ref{Q2}). 

One of the reasons we have to be interested in the Nil black hole solution is its asymptotic behavior. As we have already pointed out, in the region of large $\rho$, Nil geometry exhibits an anisotropic scale invariance similar to that of the geometries considered in the context of non-relativistic holography. Therefore, the interpretation of a black hole solution with finite temperature embedded in such spaces could presumably have some application for the holographic description of strongly correlated systems with hyperscaling \cite{hyperscaling} at finite temperature. It would be interesting if the results for the charges and thermodynamic variables derived in this paper, as well as the geometry (\ref{larotante}), had applications in that context. The holographic interpretation of black hole solutions similar to those discussed here were studied in \cite{Raulo}. 

\section*{Acknowledgments}
MO is partially funded by Beca ANID de Doctorado grant 21222264. JF is partially funded by Beca ANID de Doctorado grant 21212252. This work was partially funded by FONDECYT Regular Grants 1221504, 1200022, 1200293, 1210500, 1210635.


\begin{thebibliography}{99}

\bibitem{blackring1}
R.~Emparan and H.~S.~Reall,
``A Rotating black ring solution in five-dimensions,''
Phys. Rev. Lett. \textbf{88} (2002), 101101
[arXiv:hep-th/0110260 [hep-th]].

\bibitem{blackring2}
R.~Emparan and H.~S.~Reall,
``Black Holes in Higher Dimensions,''
Living Rev. Rel. \textbf{11} (2008), 6
[arXiv:0801.3471 [hep-th]].


\bibitem{blacksaturn}
H.~Elvang and P.~Figueras,
JHEP \textbf{05} (2007), 050
[arXiv:hep-th/0701035 [hep-th]].

\bibitem{topological}
D.~Birmingham,
``Topological black holes in Anti-de Sitter space,''
Class. Quant. Grav. \textbf{16} (1999), 1197-1205
[arXiv:hep-th/9808032 [hep-th]].

\bibitem{Nil}
C.~Cadeau and E.~Woolgar,
``New five-dimensional black holes classified by horizon geometry, and a Bianchi VI brane world,''
Class. Quant. Grav. \textbf{18} (2001), 527-542
[arXiv:gr-qc/0011029 [gr-qc]].

\bibitem{Nil2}
M.~Hassa\"\i{}ne,
``New black holes of vacuum Einstein equations with hyperscaling violation and Nil geometry horizons,''
Phys. Rev. D \textbf{91} (2015) no.8, 084054
[arXiv:1503.01716 [hep-th]].

\bibitem{Nil3}
F.~Faedo, S.~Klemm and P.~Mariotti,
``Rotating black holes with Nil or SL(2, \ensuremath{\mathbb{R}}) horizons,''
JHEP \textbf{05} (2023), 138
[arXiv:2212.04890 [hep-th]].


\bibitem{Juan}
G.~Giribet, J.~Laurnagaray and P.~Schmied,
``Probing the near-horizon geometry of black rings,''
Phys. Rev. D \textbf{108} (2023) no.2, 024061
[arXiv:2304.14461 [hep-th]].

\bibitem{Lifshitz}
S.~Kachru, X.~Liu and M.~Mulligan,
``Gravity duals of Lifshitz-like fixed points,''
Phys. Rev. D \textbf{78} (2008), 106005
[arXiv:0808.1725 [hep-th]].


\bibitem{Donnay1}
L.~Donnay, G.~Giribet, H.~A.~Gonzalez and M.~Pino,
``Supertranslations and Superrotations at the Black Hole Horizon,''
Phys. Rev. Lett. \textbf{116} (2016) no.9, 091101
[arXiv:1511.08687 [hep-th]].


\bibitem{Donnay2}
L.~Donnay, G.~Giribet, H.~A.~Gonz\'alez and M.~Pino,
``Extended Symmetries at the Black Hole Horizon,''
JHEP \textbf{09} (2016), 100
[arXiv:1607.05703 [hep-th]].

\bibitem{Strominger}
S.~W.~Hawking, M.~J.~Perry and A.~Strominger,
``Soft Hair on Black Holes,''
Phys. Rev. Lett. \textbf{116} (2016) no.23, 231301
[arXiv:1601.00921 [hep-th]].

\bibitem{Hawking}
S.~W.~Hawking,
``The Information Paradox for Black Holes,''
[arXiv:1509.01147 [hep-th]].

\bibitem{hyperscaling}
X.~Dong, S.~Harrison, S.~Kachru, G.~Torroba and H.~Wang,
``Aspects of holography for theories with hyperscaling violation,''
JHEP \textbf{06} (2012), 041
[arXiv:1201.1905 [hep-th]].

\bibitem{Raulo}
R.~E.~Arias and I.~Salazar Landea,
``Thermoelectric Transport Coefficients from Charged Solv and Nil Black Holes,''
JHEP \textbf{12} (2017), 087
[arXiv:1708.04335 [hep-th]].







\end{thebibliography}
\end{document}